\documentclass[useAMS,usenatbib]{mn2e}
\usepackage{graphicx}
\usepackage[varg]{txfonts}
\usepackage{natbib}
\title[Properties of Whiting\,1 from NIR Photometry]{Properties of the 
Young Milky Way Globular Cluster Whiting\,1 from Near-Infrared Photometry}

\author[A. T. Valcheva et al. ]{A. T. Valcheva$^{1}$\thanks{E-mail:
valcheva@phys.uni-sofia.bg}, E. P. Ovcharov$^{1}$, A. D. Lalova$^{2}$, 
P. L. Nedialkov$^{1}$, V. D. Ivanov$^{3}$ \newauthor and G. Carraro$^{3}$ \\
$^{1}$Department of Astronomy, University of Sofia, 5 James Bourchier Blvd.,
  Sofia 1164, Bulgaria\\
$^{2}$Institute of Optical Materials and Technologies, Bulgarian Academy of Sciences, 
  109 Acad. G. Bontchev Str., 1113 Sofia, Bulgaria\\
$^{3}$European Southern Observatory, Ave. Alonso de Cordova 3107, 
  Casilla 19, Santiago 19001, Chile}
\begin{document}

\date{Accepted 2014 October 9. Received 2014 October 9; in original form 2013 December 26}

\pagerange{\pageref{firstpage}--\pageref{lastpage}} \pubyear{2013}

\maketitle

\label{firstpage}

\begin{abstract}
Whiting\,1 is a member of the fast-growing group of young globular
clusters in the Milky Way halo. 
Preliminary estimates of its fundamental parameters have been provided using optical photometry and low
resolution spectroscopy.
In an attempt to strengthen our knowledge of Whiting\,1, in this study we employ a complementary approach.
Isochrone fitting method was applied on the Near-Infrared Color-Magnitude Diagram and yields an age 
t=5.7$\pm$0.3\,Gyr, metallicity $z$=0.006$\pm$0.001 ([Fe/H]=$-$0.5$\pm$0.1) 
and distance modulus $(m-M)_0$=17.48$\pm$0.10. Our results confirm that Whiting\,1 is a young and moderately metal-rich 
globular cluster. It is one of the youngest from the Sgr dSph. We fitted an Elson, Fall and Freeman (EFF)
profile to the near-infrared number counts, and measured cluster core radius 
$r_c$=9.1$\arcsec$$\pm$3.9$\arcsec$.
Two probable eclipsing variables in the cluster were found from
multi-epoch $V$ band photometry. Finally, an unknown galaxy cluster was identified on our $K$ vs. $(J-K)$ color-magnitude diagram.
It has a redshift z$\sim$1, and it is located at about 1$\arcmin$ from the center of Whiting\,1
at $\alpha_{J2000}=02^{h} 02^{m} 56.6^{s}$, $\delta_{J2000}=-03^{\circ} 16\arcmin 09\arcsec$, 
contaminating the cluster photometry.
\end{abstract}


\section{Introduction}

Whiting\,1 ($\alpha_{J2000}=02^{h} 02^{m} 57^{s}$, $\delta_{J2000}=-03^{\circ} 15\arcmin 10\arcsec$) 
is a halo globular cluster at a heliocentric distance D$\sim$45\,kpc, as first shown 
by \citet{car05}. It was discovered by \citet{whi02} during a search 
for Galaxies in the Zone of Avoidance, and originally it was classified as an 
open cluster by \citet{dia02}. Later, \citet{car07} 
confirmed its association with the Milky Way halo with deep $BVI$ 
photometry, and updated the heliocentric distance to 
D$\sim$29.4\,kpc. They derived an age of 6.5\,Gyr and metallicity 
$z$=0.004$\pm$0.001 from isochrone fitting and estimated A$_{V}$=0.11 mag and (m-M)$_{0}$=17.34. 
In latter study, measured CaII indexes ($\Sigma$CaII) of
three stars in Whiting 1 give metallicity [Fe/H] ranging from $-$1.1 to $-$0.4 
($z$ ranging from 0.002 to 0.008).
They pointed out that Whiting\,1 is 
spatially close to the trailing stream of Sgr dSph, and its kinematics is 
consistent with that of this galaxy. Besides, it follows very nicely the Sgr dSph 
age-metallicity relation \citep{forb04, siegel}. 

Only a few globular clusters in the halo of the Galaxy are known to 
have comparably young ages ($i.e.$, 
Pal\,1 -- Rosenberg et al. 1998a;  
Pal\,12 -- Rosenberg et al. 1998b; 
Segue\,3 -- Ortolani et al. 2013; Crater -- Belokurov et al. 2014), 
making Whiting\,1 a member of a rare class of objects.

Up to now no NIR photometry of Whiting\,1 has been 
reported in the literature and in this study we aim to obtain the fundamental properties through the NIR CMD 
using data from the ESO Science Archive\footnote{http://archive.eso.org}. Also, a search for variable star population 
has never been performed.

The next section describes our data. Section 3, 4 and 5 present our analysis and results of the NIR data: 
cluster structural parameters and cluster fundamental parameters - age, metallicity and distance modulus. Section 6 
describes our search for variable stars in $V$ band and our conclusions are drawn in Section 7.

\section[]{Observations and Data Reduction}
\subsection{NIR observations}
Whiting\,1 was observed in the NIR on 2006 September 4 at the ESO 
NTT (New Technology Telescope) with SofI 
\citep[Son of ISAAC;][]{1998Msngr..91....9M} in 
large-field imaging mode. SofI is equipped with a 1024$\times$1024 
detector, with a pixel scale of 0.292\,arcsec\,pix$^{-1}$ and FOV of 
4\arcmin.98 x 4\arcmin.98. We had 
15 images in $J_S$ (each was the average of 4 frames of 15\,sec 
integration) and 32 images in $K_S$ (each was the average of 6 
frames of 10\,sec integration).  The seeing during the 
observations was 0\arcsec.8 -- 0\arcsec.9 on average for both 
filters. The data reduction was carried out 
with IRAF\footnote{Image Reduction and Analysis Facility is a general 
purpose software system for the reduction and analysis of astronomical 
data, IRAF is written and supported by the IRAF programming group at 
the National Optical Astronomy Observatories (NOAO).}, and included 
flat fielding, sky/bias/dark subtraction, 
image alignment and combination. The sky frame was created by 
median combination of all images with the task {\it imcombine},
applying the {\it avsigclip} 3-$\sigma$ rejection algorithm, and 
appropriate thresholds to excise the stars. A constant was added to 
the sky subtracted images during the final combination to set their
median value to zero, accounting for the sky variations during the 
observations. Next, we performed 
PSF photometry (Stetson 1987) with DAOPHOT/IRAF. Only 99 stars were measured in both $J_S$ and 
$K_S$. The 100\% completeness limits were defined from the first drop in the luminosity function 
which give $J_S^{lim}\sim$19.5\,mag and 
$K_S^{lim}\sim$18.2\,mag (magnitude bins varied from 0.1 to 0.3), and the photometric errors for 
stars brighter than these limits are typically less than 0.1\,mag. 
The photometric calibration was based on 9 stars for $J_S$, and 7 
stars for $K_S$, in common with the Two Micron All Sky Survey \citep[2MASS]{skrut}. 
\begin{equation}
J = j_s-2.208 (\pm0.019),
\end{equation}
\begin{equation}
K_S = k_s-2.894 (\pm0.023),
\end{equation}
where the lower case letters mark the instrumental magnitudes, and 
the upper case ones mark the standard magnitudes (hereafter we will use 
the designation K for $K_{S}$ filter).

\subsection{Optical Variability Monitoring}

Optical images of Whiting\,1 were collected with EMMI (ESO 
Multi-Mode Instrument; D'Odorico 1990) and EFOSC2 (ESO Faint Object 
Spectrograph and Camera; Snodgrass et al. 2008) on the 3.58-m ESO 
NTT (New Technology Telescope) at La Silla, Chile, typically to 
fill small gaps in other programs or during periods of poor weather 
conditions. Binning by a 
factor of two was used on all occasions, making the pixels scales 
0.33 and 0.24\,arcsec\,pix$^{-1}$ and the FOV 5\arcmin.6 x 9\arcmin.9 and 
4\arcmin.1 x 4\arcmin.1, respectively for both 
instruments. Five epochs were obtained in $V$ band, and the first 
was paired with $I$ band for quasi-simultaneous color information. 
The total integration for each epoch was split into three images, 
taken with some offsets to minimize the effects from the detector's 
cosmetics. A detailed log of the observations is presented in 
Table~\ref{Table_ObsLog}. 
\begin{table}
\caption{Log of the optical monitoring observations.}\label{Table_ObsLog}
\centering
\begin{tabular}{@{ }c@{ }c@{ }c@{ }c@{ }c@{ }c@{ }c@{ }c@{ }}
\hline
Epoch & Date       & UTC      &~Exp.~&~Instru-~&~Fil-~&~FWHM~    \\
No.   &~yyyy/mm/dd~&~hh:mm:ss~& [sec]& ment    & ter  & [arcsec] \\
\hline
1     & 2006/11/11 & 00:49:07 & 200  & EMMI    & V    & 1.0      \\
1     & 2006/11/11 & 00:53:00 & 200  & EMMI    & V    & 0.85     \\
1     & 2006/11/11 & 00:56:54 & 200  & EMMI    & V    & 1.0      \\
1     & 2006/11/11 & 01:03:06 & 200  & EMMI    & I    & 1.0      \\
1     & 2006/11/11 & 01:00:00 & 200  & EMMI    & I    & 1.0      \\
1     & 2006/11/11 & 01:00:00 & 200  & EMMI    & I    & 1.0      \\
\hline
2     & 2006/11/11 & 05:43:03 & 200  & EMMI    & V    & 0.85     \\
2     & 2006/11/11 & 05:46:56 & 200  & EMMI    & V    & 0.83     \\
2     & 2006/11/11 & 05:50:51 & 200  & EMMI    & V    & 0.83     \\
\hline
3     & 2006/11/29 & 05:42:11 & 200  & EFOSC   & V    & 1.3      \\
3     & 2006/11/29 & 05:46:18 & 200  & EFOSC   & V    & 1.3      \\
3     & 2006/11/29 & 05:50:26 & 200  & EFOSC   & V    & 1.3      \\
\hline
4     & 2006/12/18 & 02:52:19 & 200  & EFOSC   & V    & 0.8      \\
4     & 2006/12/18 & 02:56:19 & 200  & EFOSC   & V    & 0.8      \\
4     & 2006/12/18 & 03:00:19 & 200  & EFOSC   & V    & 0.8      \\
\hline
5     & 2007/01/27 & 01:18:40 & 200  & EMMI    & V    & 1.65     \\
5     & 2007/01/27 & 01:23:10 & 200  & EMMI    & V    & 1.65     \\
5     & 2007/01/27 & 01:27:45 & 200  & EMMI    & V    & 1.65     \\
\hline
\end{tabular}
\end{table}
The basic reduction was carried out with IRAF. For every epoch, we flat-fielded, 
aligned and co-added all images in the same filter. Next, we performed 
PSF photometry. The astrometric calibration is based on 
ten stars from Guide Star Catalog 2.3.2 \citep{las08} and the 
photometric calibration is based on 210 stars we have in common with the 
catalog of \citet{car07}. The derived photometric 
transformation equations are:  
\begin{equation}
V_{1} = v_{1}+1.921 (\pm0.012),
\end{equation}
\begin{equation}
V_{2} = v_{2}+1.849 (\pm0.012),
\end{equation}
\begin{equation}
V_{3} = v_{3}+2.222 (\pm0.016),
\end{equation}
\begin{equation}
V_{4} = v_{4}+2.047 (\pm0.019),
\end{equation}
\begin{equation}
V_{5} = v_{5}+1.658 (\pm0.015), 
\end{equation}
\begin{equation}
I_{1} = -0.09(\pm0.20)+1.11(\pm0.01)*i_{1}, for\,i_{1} \le 17\,mag,
\end{equation}
\begin{equation}
I_{1} = i_{1}+1.737 (\pm0.005), for\,i_{1} > 17\,mag, 
\end{equation}
where the lower case letters mark the instrumental magnitudes, and 
the upper case ones mark the standard magnitudes. The subscripts give 
the epoch number, as defined in Table\,\ref{Table_ObsLog}. No 
significant color terms were found. 

We cross-identified all photometry lists. For the first epoch 865 stars were 
measured in $V$ and $I$ bands simultaneously. There are 273 stars with three $V$ band 
epochs, 259 star with four, and 231 with five. The EFOSC2's images are nearly two times smaller 
compared to EMMI's and the cluster is not centered 
in the EMMI's images. This highly reduces the number of common stars.

An excerpt of the photometric catalog is given in Table\,\ref{tab:phot_all}.
\begin{table*}
\caption{Photometry of the stars in Whiting\,1 and the surrounding 
field. The complete catalogue is available online.}\label{tab:phot_all}
\begin{tabular}{@{}c@{ }c@{ }c@{ }c@{ }c@{ }c@{ }c@{ }c@{ }c@{ }c@{ }c@{ }c@{ }c@{ }c@{ }c@{ }c@{ }c@{ }c@{ }c@{}}
\hline
N & $\alpha_{J2000}$ & $\delta_{J2000}$ & $V$$_1$ & ~$\sigma$($V$$_1$)~ & $V$$_2$ & ~$\sigma$($V$$_2$)~ & $V$$_3$ & ~$\sigma$($V$$_3$)~ & $V$$_4$ & ~$\sigma$($V$$_4$)~ & $V$$_5$ & ~$\sigma$($V$$_5$)~ & $I$$_1$ & ~$\sigma$($I$$_1$)~ & $J$ & ~$\sigma$($J$)~ & $K_S$ & ~$\sigma$($K_S$) \\
  & hh:mm:ss        & dd:mm:ss        & mag     & mag                 & mag     & mag                 & mag     & mag                 & mag     & mag                 & mag     & mag                 & mag     & mag                 & mag & mag             & mag   & mag              \\
\hline
 1 &~2:02:54.63~& $-$3:12:43.8~&~23.336~& 0.043 & 23.491 & 0.082 & 23.175 & 0.086 & 23.378 & 0.052 & 23.386 & 0.241 & 21.629 & 0.031 & 20.312 & 0.100 & 18.804 & 0.164  \\
 2 &~2:03:03.82~& $-$3:12:52.1~& 22.434 & 0.027 & 22.433 & 0.038 & 22.706 & 0.059 & 22.577 & 0.039 & 22.261 & 0.080 & 21.518 & 0.030 & 19.653 & 0.082 & 18.705 & 0.170 \\
 3 &~2:02:59.28~& $-$3:13:07.8~& 21.731 & 0.121 & 21.709 & 0.017 & 21.844 & 0.030 & 21.771 & 0.014 & 21.754 & 0.056 & 20.458 & 0.025 & 19.722 & 0.060 & 18.835 & 0.164  \\
 4 &~2:02:54.22~& $-$3:13:07.6~& 22.045 & 0.046 &        &       &        &       &        &       &        &       & 20.229 & 0.036 & 19.171 & 0.074 & 17.641 & 0.069 \\
 5 &~2:02:57.19~& $-$3:13:12.2~& 22.847 & 0.087 & 22.850 & 0.082 & 22.524 & 0.054 & 22.651 & 0.065 & 22.525 & 0.092 & 20.843 & 0.073 & 19.566 & 0.093 & 18.049 & 0.129 \\
 6 &~2:03:00.11~& $-$3:13:24.9~& 23.454 & 0.082 & 23.405 & 0.093 & 23.233 & 0.086 & 23.346 & 0.076 & 22.925 & 0.173 & 21.840 & 0.099 & 20.003 & 0.100 & 17.798 & 0.102 \\
 7 &~2:03:02.95~& $-$3:13:31.9~& 23.004 & 0.035 & 23.001 & 0.054 & 23.036 & 0.070 & 23.031 & 0.053 & 22.785 & 0.148 & 21.268 & 0.018 & 20.095 & 0.088 & 18.565 & 0.126 \\
 8 &~2:03:02.26~& $-$3:13:42.1~& 22.604 & 0.033 & 22.560 & 0.037 & 22.589 & 0.050 & 22.586 & 0.037 & 22.481 & 0.096 & 20.635 & 0.034 & 19.487 & 0.060 & 17.934 & 0.090 \\
 9 &~2:03:03.42~& $-$3:13:45.1~& 21.370 & 0.053 & 21.358 & 0.050 & 21.349 & 0.040 & 21.358 & 0.041 & 21.201 & 0.039 & 20.420 & 0.050 & 19.705 & 0.092 & 18.456 & 0.141  \\
10 &~2:03:03.27~& $-$3:13:50.9~& 22.804 & 0.024 &        &       &        &       &        &       &        &       & 20.632 & 0.033 & 19.500 & 0.053 & 18.800 & 0.157 \\
\hline
\end{tabular}
\end{table*}
\section{Cluster Radius and Structural parameters from the NIR 
Observations}\label{sec:rp}

The wide field of view of our NIR images (4.92$\arcmin$$\times$4.92$\arcmin$) 
allows us to verify the cluster radius and structural parameters. 
The previous radius estimates range from 0\arcmin.5, obtained from the radial density 
distribution of stars \citep{car05},  
to 0\arcmin.6 - angular apparent radius \citep{dia02}. The deeper photometry of \citet{car07} suggests, 
based on a simple visual inspection, that the cluster may extent out to 1\arcmin.5. \citet{car07} fitted King's 
profile \citep{king62} to obtain a core radius $r_c$=15$\arcsec$. This 
work showed an excess of stars in the outer region, approximated by 
a power law, attributed to tidal stripping.

\begin{figure}
 \resizebox{\hsize}{!}{\includegraphics{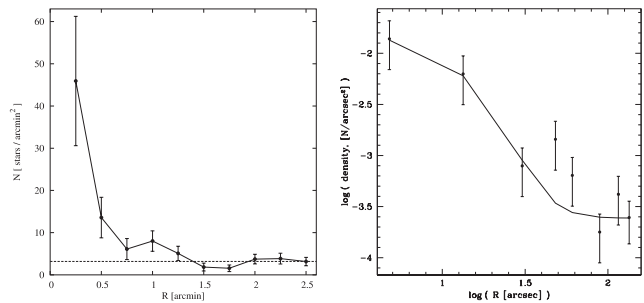}}
\caption{\textbf{Left:} Radial density distribution of all stars in the field 
in concentric rings around the cluster center with 0.25$\arcmin$ bin. Contamination level, 
estimated at distance 1.75$\arcmin$ $<$ r $<$ 2.5$\arcmin$, is shown with dashed line. Error bars give Poisson uncertainties. 
\textbf{Right:} Elson, Fall \& Freeman profile \citep{Els87} fit to the observed 
radial profile. See Sect.\,\ref{sec:rp} for the fitting parameters.}
\label{rad}
\end{figure}

We built a radial profile counting stars within 0.25$\arcmin$ wide 
concentric annuli, centered on Whiting\,1, and normalized them to 
unit area (Fig.\,\ref{rad}, left). Throughout our analysis we used cluster center coordinates from 
\citet{dia02}, cited in the introduction of this paper. The profile shows some excess 
counts over the background at r$\sim$1$\arcmin$-1.25$\arcmin$ but we 
conservatively adopt a cluster radius of 0.75$\arcmin$ for the 
following Color-Magnitude Diagrams (CMD) analysis, to minimize 
the field contamination. 
Our attempts to fit King's profile failed, 
possibly due to the small number of stars. However, we could fit 
Elson, Fall \& Freeman's profile (hereafter EFF; Elson et al. 1987; Fig.\,\ref{rad}, right): 
\begin{equation}
n(r)=n_0\times(1+r^2/a^2)^{-\gamma/2}+\Phi. 
\end{equation}
where $n_0$ is central star-density, $\gamma$ is power-law slope,
$\Phi$ is the background contamination level, $a$ is a parameter, 
related to the core radius $r_c$ via 
$r_c$=$a$$\times$(2$^{2/\gamma}$$-$1)$^{1/2}$.
We only considered stars with $K_0$$\leq$18.6\,mag and 
($J-K$)$_0$$\leq$0.99\,mag to remove the bulk of background 
galaxies, and obtained: 
$n_0$=0.0153$\pm$0.0064\,stars/$\square$$\arcsec$,
$\gamma$=5.27$\pm$1.53, 
$\Phi$=0.000244$\pm$0.000083\,stars/$\square$$\arcsec$,
a=20.0$\arcsec$$\pm$4.0$\arcsec$, r$_c$=9.09$\arcsec$$\pm$3.68$\arcsec$. 
Our NIR core radius is similar to the optical value of \citet{car07}.

\section{NIR Color-Magnitude and Color-Color Diagrams}

\begin{figure}
\resizebox{\hsize}{!}{\includegraphics{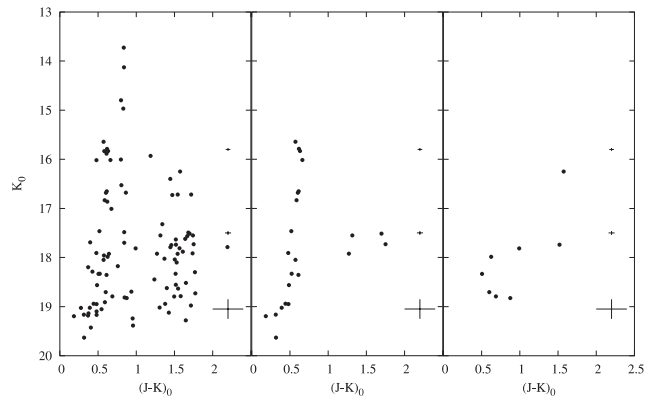}}
\caption{CMDs for all stars in the observed field (left), for the 
``cluster'' stars within 0.75$\arcmin$ from the cluster center 
(middle), and for stars in a control ring at r=2$\arcmin$ with an 
area equal to the ``cluster'' area. Typical errors for stars  with ($J$$-$$K$$_{0}$) $\sim$ 0.5 mag 
are shown on the right. Extinction correction for E($J$$-$$K$)=0.013\,mag \citep{sch98} 
is applied to all stars.}\label{cmd}
\end{figure}

CMDs for Whiting\,1 and the surrounding field are shown in Fig.\,\ref{cmd}, and a comparison of probable 
cluster members, defined to lie within 0.75$\arcmin$ from the cluster center with an equal area background field.

The cluster members occupy a locus at ($J$$-$$K$)$_0$$\sim$0.3--0.7\,mag and 
$K_{0}$$\sim$15.5--19.5\,mag (Fig. \ref{cmd}, center). The faintest cluster stars we 
detected are sub-giants at $K$$_0$$\sim$18.7--19.5\,mag. The extremely red objects at 
($J$$-$$K$)$_0$$\sim$1.3--1.7\,mag and $K_{0}$$\sim$17.5--18\,mag probably belong to the field, as 
suggested by the right panel in Fig.\,\ref{cmd} and by the Besan\c con model of stellar population synthesis 
of the Galaxy \citep{rob03}, which predicts only 2 foreground stars between 15 mag and 18 mag in $K$ band in the 
cluster's field (r=0\arcmin.75). The expected number of stars for the whole imaged field predicted by Besan\c con model 
is 25 stars in the magnitudes interval [13-20] mag.

\begin{figure}
\resizebox{\hsize}{!}{\includegraphics{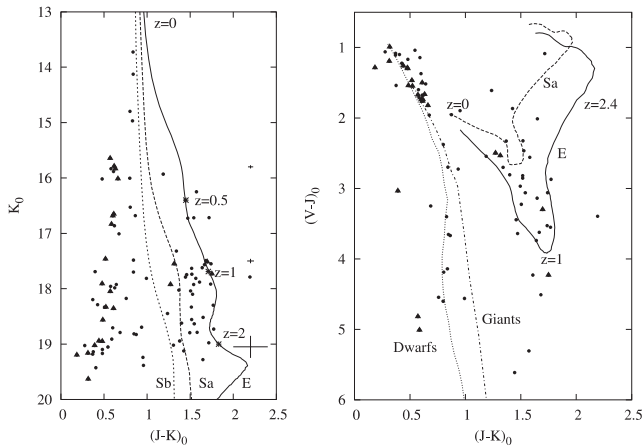}}
\caption{{\bf Left}: Dereddened NIR CMD of all sources in the field 
(4.92$\arcmin$$\times$ 4.92$\arcmin$). The lines are GALEV 
models \citep{kot09} for E (solid), Sa (dashed) and Sb 
(dotted) type galaxies, over cosmological timescales. Redshifts z 
are marked. Typical error bars for stars  with ($J$$-$$K$$_{0}$) $\sim$ 0.5 mag are shown on 
the right.
{\bf Right}: Dereddened optical-NIR color-color diagram. GALEV 
models (Kotulla et al. 2009) for E (solid line) and Sa (dashed line) 
type galaxies, are shown, together with dwarfs (dotted line) and giant 
(dot-dashed line) star sequences from \citet{fro78}.
{\bf In both panels}: Candidate member stars (within 
0.75$\arcmin$ from the cluster center) are marked with solid 
triangles, other sources are plotted with solid dots.}
\label{discuss}
\end{figure}

The CMD of all stars in the field (Fig.\,\ref{cmd}, left) has an 
abundance of sources with ($J$$-$$K$)$_0$$\ge$1.2\,mag, possibly
distant galaxies (Fig.\,\ref{discuss}). Indeed, their colors are
consistent with those of GALEV evolutionary synthesis models \citep{kot09} for E/Sa 
galaxies at z$\sim$1 (Fig.\,\ref{discuss}, 
left). The spatial distribution of the sources showed some 
clustering of sources with $J$=19--20\,mag and 
$(J-K)_0$$\sim$1.6\,mag within a circle with $\sim$1.5$\arcmin$ 
diameter at $\alpha_{J2000}=02^{h}02^{m}56.6^{s}$, $\delta_{J2000}=-03^{\circ} 16\arcmin 09\arcsec$.
This may be an unknown galaxy cluster: the NASA Extragalactic 
database\footnote{This research has made use of the NASA/IPAC 
Extragalactic Database (NED) which is operated by the Jet Propulsion 
Laboratory, California Institute of Technology, under contract with 
the NASA.} 
identified four galaxies with unknown redshift within 2$\arcmin$ 
from Whiting\,1; the nearest galaxy cluster reported by \citet{wen12} in their catalog of SDSS-III clusters is 
$\sim$10.5$\arcmin$ away, and it has photometric redshift 
z$\sim$0.49. Finally, the bluer field objects with 
($J$$-$$K$)$_0$$\le$1.2\,mag appears to be dwarf stars, as seen
from the right panels in Fig.\,\ref{discuss} where we 
compare their location with the colors of dwarfs and giant stars \citep{fro78}.

\section{Fundamental properties via isochrone fitting}

We fitted the CMD of stars within 0.75$^\prime$ radius from 
the Whiting\,1 center with Padova PARSEC v1.1 isochrones \citep{bress}. 
We varied the metallicity $z$ from 0.001 to 0.01 with a step of 0.0001 
(91 different values) and the age t from 4.77 Gyr to 7.94 Gyr with a 
step $\Delta$(log\textit{t})=0.001 ($\sim$10 Myr) (221 different values). 
Constant extinction of E($J$$-$$K$)=0.013\,mag \citep{sch98} was applied 
to all stars. We kept the reddening constant because it is not expected 
to vary significantly at that Galactic latitude, and it has a small 
effect in the NIR.
\begin{figure*}
\centering\includegraphics[width = 0.438\textwidth]{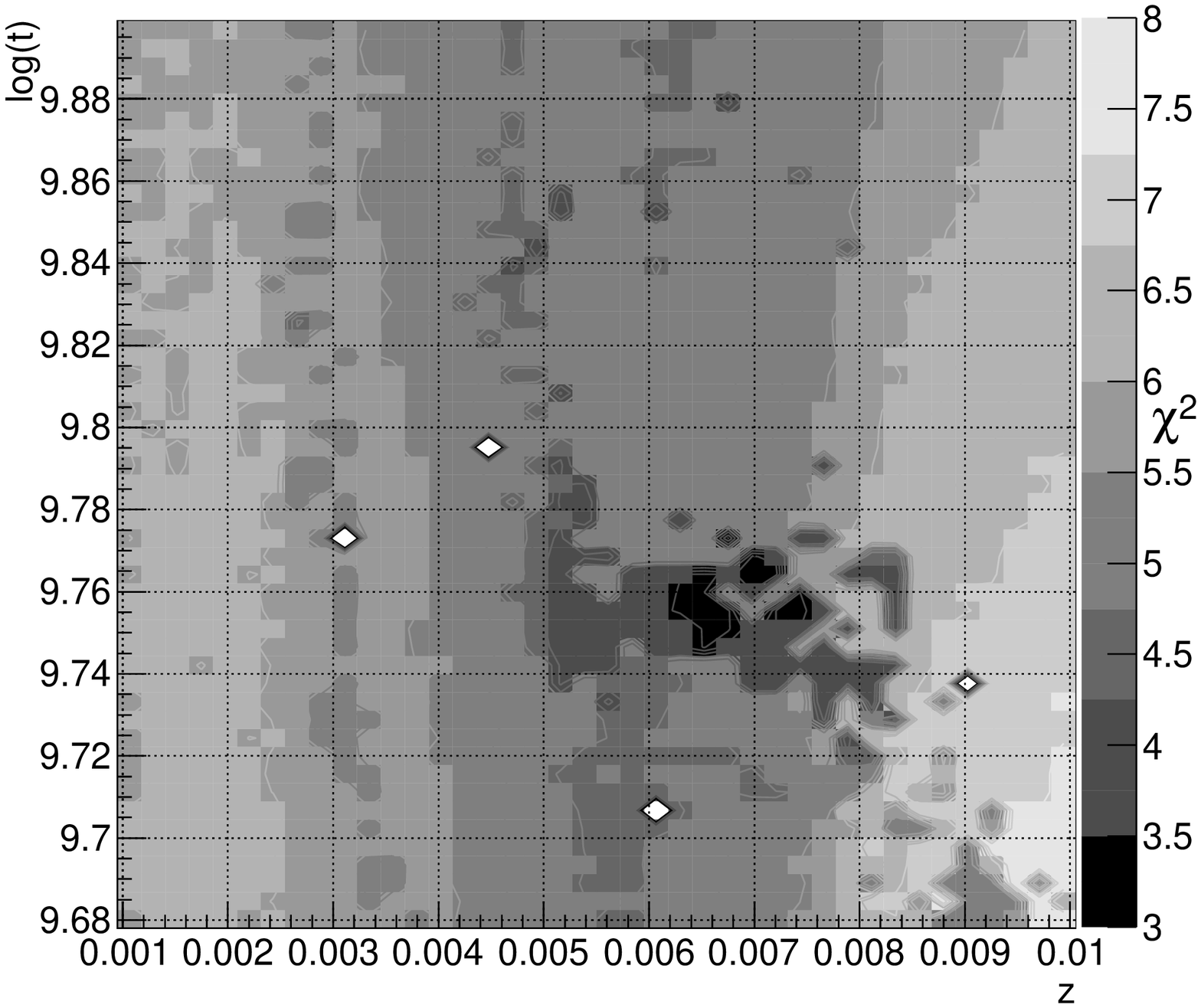}
\centering\includegraphics[width = 0.424\textwidth]{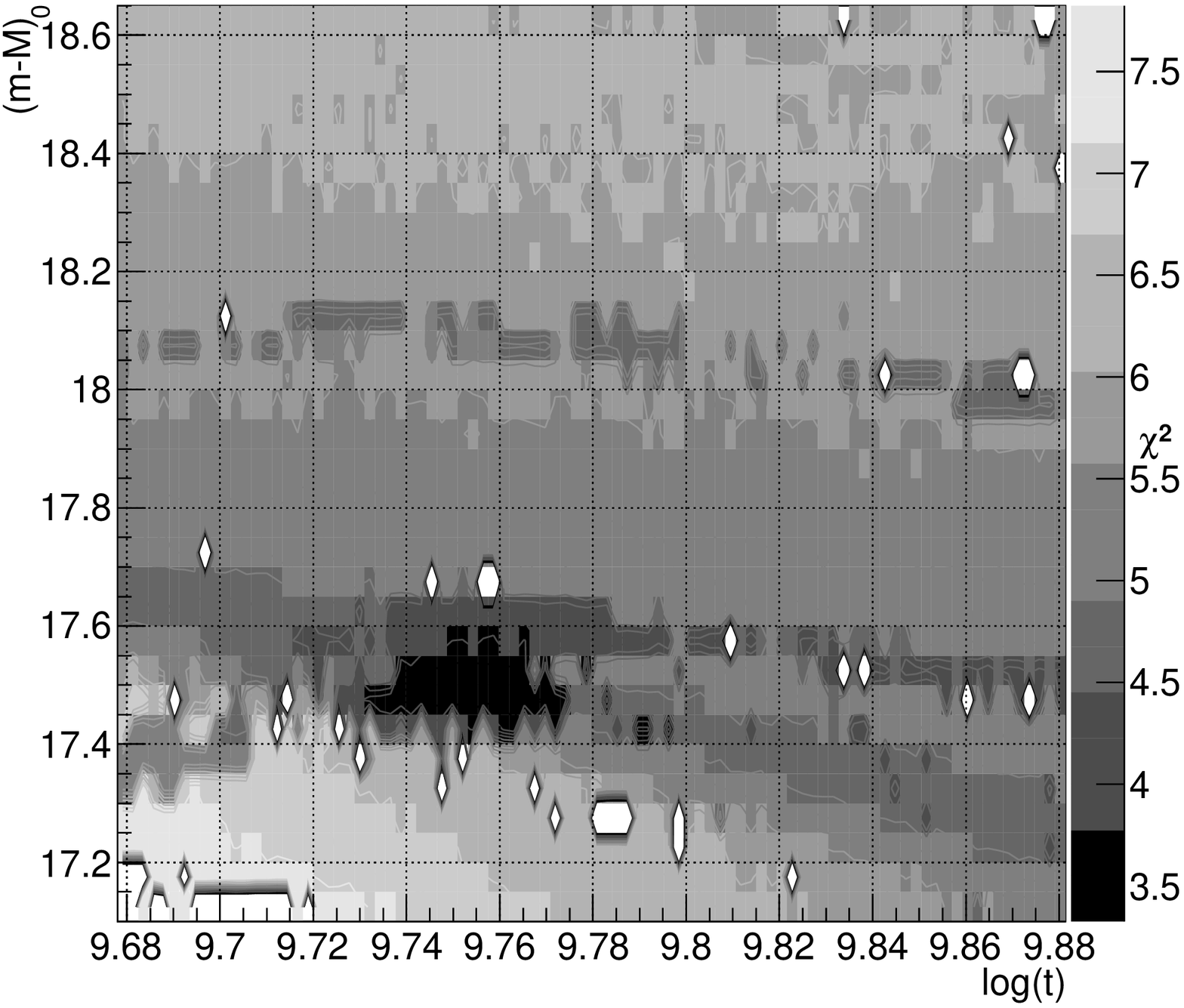}

\caption{$\chi^2$ as a function of: the 
metallicity $z$ and the age t (left panel) and t and the distance modulus (m$-$M)$_0$ (right panel).
The white blocks show unsuccessful fitting.}\label{iso_fit}
\end{figure*}

We minimized $\chi^2$ for each ($z$,t) pair treating the distance 
modulus (m$-$M)$_0$ as a free parameter. The resulting $\chi^2$ map is 
shown in Fig. \ref{iso_fit}, left panel. The relatively low values of the $\chi^2$ 
($\sim$3 near the minimum) compared to the number of the degrees of 
freedom (17) indicates an overestimation of the photometric errors. 
A parabolic fit was applied to the minimal values of $\chi^2$ as a function of t and 
the minimum suggested a preferred value of t=5.7$\pm$0.3\,Gyr. For the 
obtained t, the $\chi^2$ as a function of the metallicity gives 
z=0.0063$\pm$0.0011, again after fitting a parabola. 

Finally, the best fitting distance modulus was (m$-$M)$_0$= 
17.48$\pm$0.10. The error includes both parametric (from the variation 
of $z$ and t within their errors) and statistical (from the $\chi^2$ fit) 
error. The $\chi^2$ as a function of the age t and the distance modulus (m$-$M)$_0$ is shown in Fig. \ref{iso_fit}, right panel.


Our results are based on RGB and sub-giant stars because 
the NIR data do not reach the main sequence turn off point. The derived 
properties are consistent, within 1$\sigma$, with the ones obtained in 
the optics \citep{car07}, despite the smaller number of stars on the NIR CMD 
than on the optical CMD.

\section{Search for Variable Stars}

\begin{figure}
\resizebox{\hsize}{!}{\includegraphics{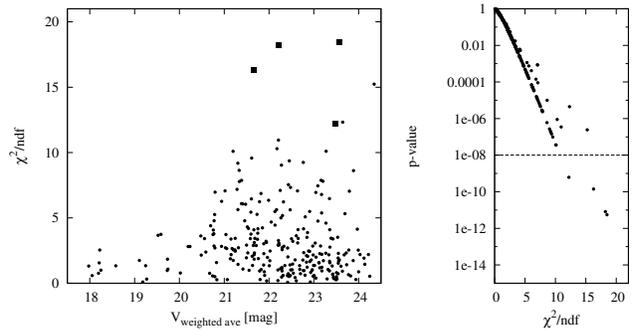}}
\caption{ Diagrams $\chi^2_n$ vs. $V_{weighted~ave}$ (left) and p-value vs. $\chi^2_n$ (right) for all 
stars with three and more epochs in $V$ band. P-values smaller than $10^{-8}$ are 
adopted as a criteria for selection of candidate variables. The four candidates are marked with solid squares in the left panel.}
\label{var_ind}
\end{figure}

We used our multi-epoch $V$ band observations to search for variable stars in Whiting\,1 and the surrounding field. 
We considered only the stars with three and more epochs - 273 stars in total. Because of the small number of measurements 
and different data quality for every star, we used normalized $\chi^2$ and p-values (the probability to obtain these data if 
the available magnitudes for every star are randomly distributed around the mean value) to select candidates for variable stars. 
Diagram $\chi^2_n$ vs. $V$ is shown on the left panel in Fig.\,\ref{var_ind}. Large dispersion is visible and expected, 
but some stars show extremely high values of $\chi^2_n$ and can be potential variable candidates ($\chi^2_n > 12$). To check 
the significance of this selection, we calculated the p-value of every $\chi^2_n$ (Fig.\,\ref{var_ind}, right panel). 
The three visible sequences are stars with 3, 4 and 5 epochs. We considered as candidates variables only those with p-value smaller 
than $10^{-8}$, where we have something like a discontinuity in the diagram. The four selected stars are also marked on the 
left panel with solid squares.  
\begin{figure}
\resizebox{\hsize}{!}{\includegraphics[width=8cm]{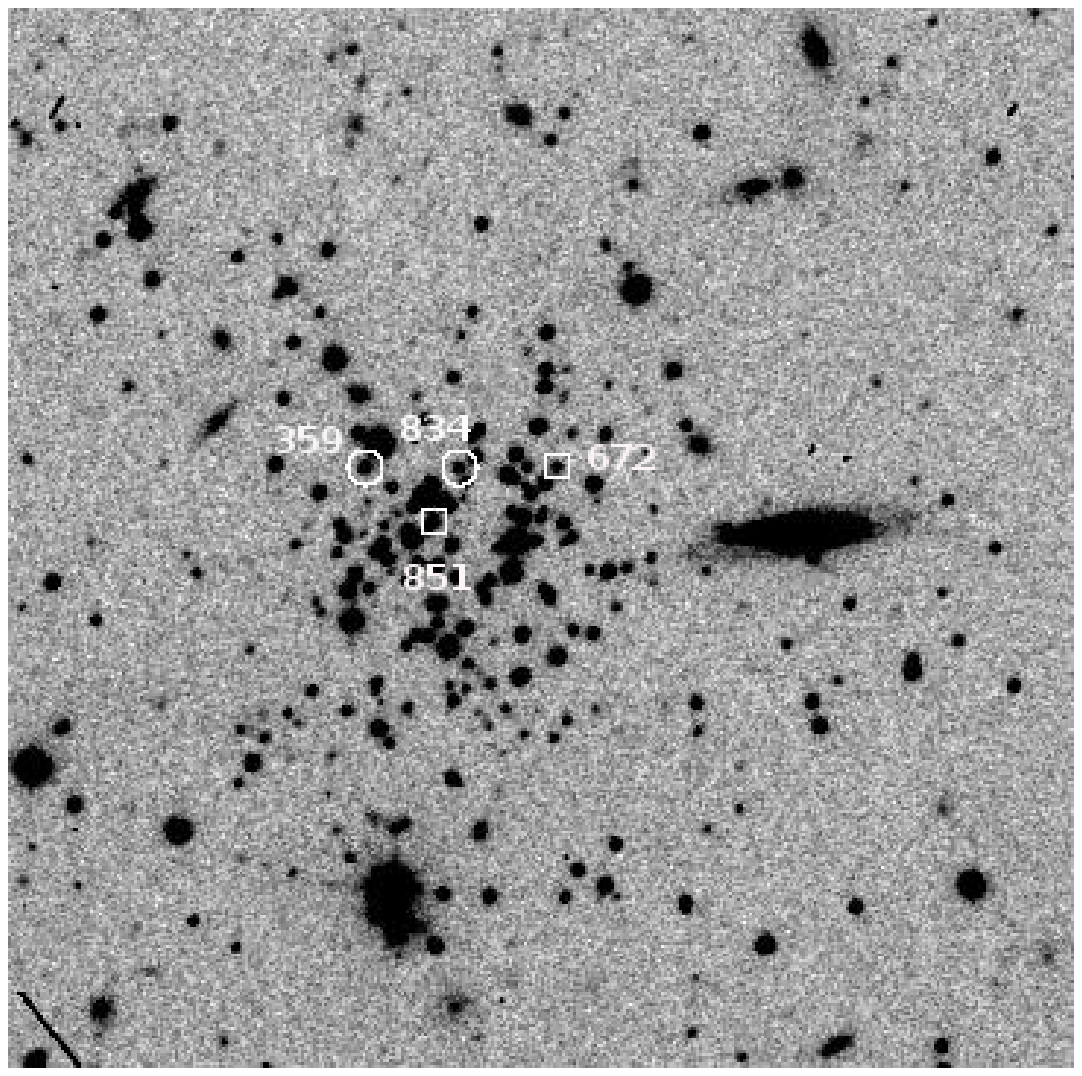}}
\caption{$V$ band image of Whiting\,1 (600 sec. of integration, 2$\arcmin$x2$\arcmin$ FoV). Candidate variables are marked with circles 
and reference stars with squares. All stars are labeled with the corresponding number in our photometric catalog. North is up and 
East is to the left. 
}\label{2var}
\end{figure}

After careful examination of the four selected candidates, we found that: the magnitude of the star \#491 (the number in our photometric catalog) 
is affected by a cosmic ray; the star \#363 is most probably the galaxy SDSS J020307.30-031501.0, which is at a distance of 
0.95$\arcsec$ away from \#363; the remaining two stars - \#359 and \#834 - exhibit statistically significant variability 
and both lie within the adopted cluster limits. They are marked in Fig.\,\ref{2var} with circles and their properties are listed in 
Table\,\ref{tab:variables_list}. 
\begin{table*}
\caption{Candidate variable stars. Available $V$ band epochs, $I$ band, errors, computed $\chi^2_n$ and p-values are listed.}
\label{tab:variables_list}
\centering
\begin{tabular}{@{}c@{ }c@{ }c@{ }c@{ }c@{ }c@{ }c@{ }c@{ }c@{ }c@{ }c@{ }c@{ }c@{ }c@{ }c@{ }c@{}c@{}c@{}}
\hline
~~N~~ & $\alpha_{J2000}$ & $\delta_{J2000}$ & $V$$_1$ & ~$\sigma$($V$$_1$)~~ & $V$$_2$ & ~$\sigma$($V$$_2$)~~ & $V$$_3$ & ~$\sigma$($V$$_3$)~~ & $V$$_4$ & ~$\sigma$($V$$_4$)~~ & $I$ & ~$\sigma$($I$)~~~~ & $\chi^2_n$ & p-value\\
      & deg             & deg             & mag     & mag                  & mag     & mag                  & mag     & mag                  & mag     & mag                  & mag     & mag                    &    &   \\
\hline
359 & 2:02:57.47 & $-$3:14:59.2 & 21.540 & 0.017 & 21.518 & 0.022 & 21.751 & 0.031 & 21.685 & 0.018 & 20.865 & 0.034 & 16.281 &~~1.408192e-10\\
834 & 2:02:56.75 & $-$3:14:59.2 & 22.151 & 0.018 & 22.160 & 0.027 & 22.090 & 0.054 & 22.343 & 0.020 & 21.332 & 0.032 & 18.218 &~~8.141098e-12 \\
\hline
\end{tabular}
\end{table*}

\begin{figure}
\resizebox{\hsize}{!}{\includegraphics{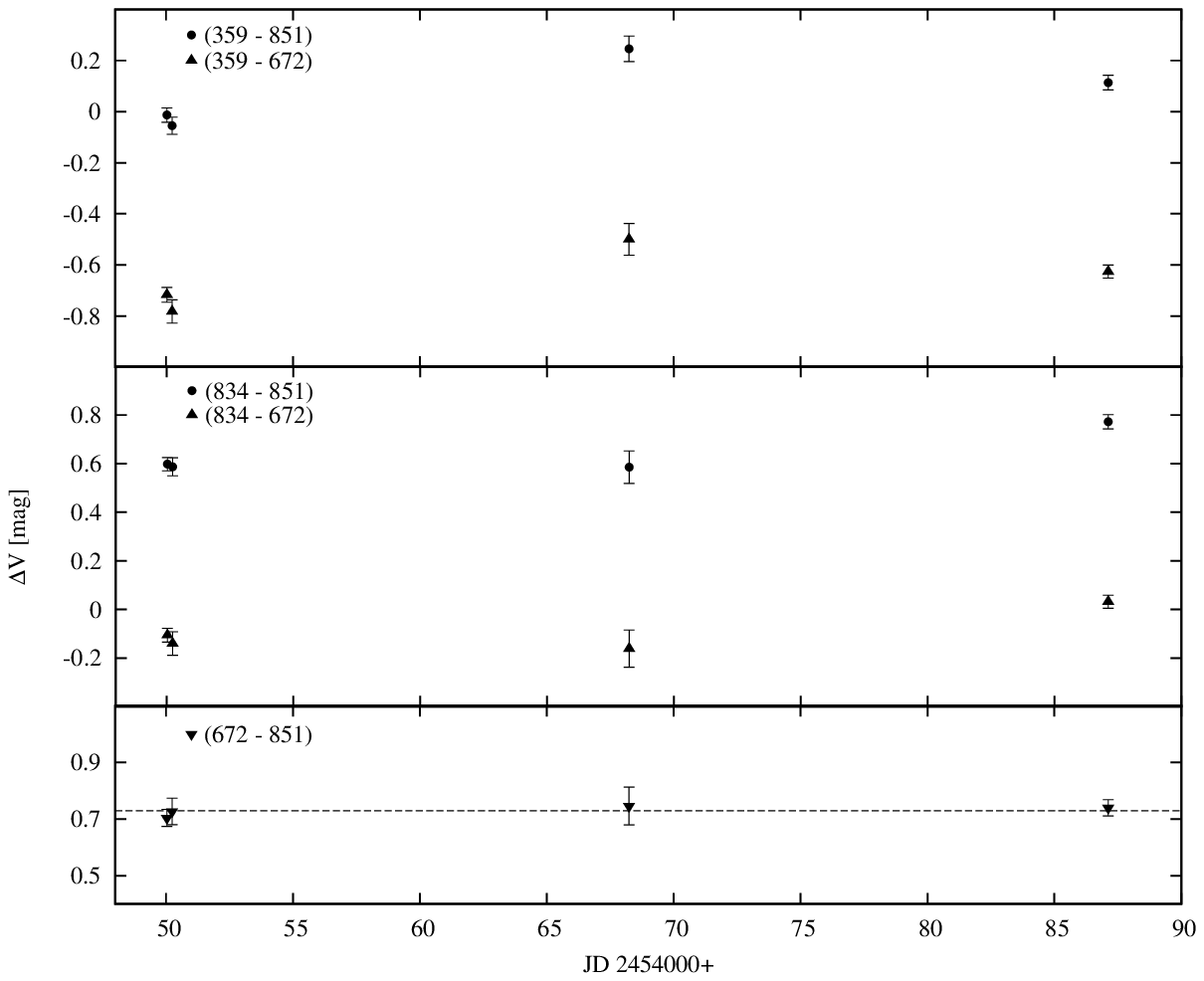}}
\caption{Differential light curves of the 2 candidate variables \#359 and \#834 
and the reference stars \#672 and \#851 are presented on the top and middle panels. The
difference between the reference stars \#672 and \#851 is shown on the bottom panel. Weighted average of the 
differences is plotted with dashed line.
}\label{light_curves}
\end{figure}

The differential light curves (Fig.\,\ref{light_curves}) of the candidate variables \#359 and \#834 were generated with respect 
to the reference stars \#672 and \#851 which are the nearest stable stars similar in magnitude and color. The maximum peak-to-peak
variation of both candidates is $\sim$ 0.2 mag over the monitoring period of 35 days with available data for 4 epochs.
\begin{figure}
\resizebox{\hsize}{!}{\includegraphics{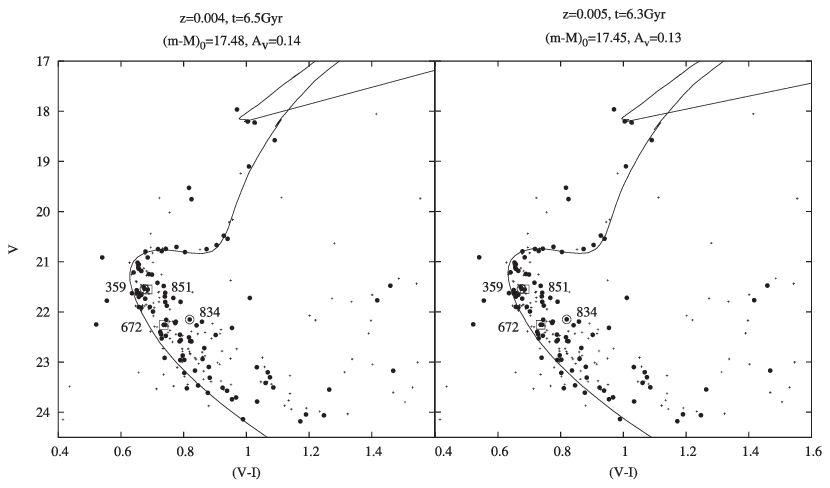}}
\caption{Optical color-magnitude diagram for all stars with 3 and more epoch in $V$ band (crosses). Stars lying within 
the determined radius of 0.75$\arcmin$ are shown with solid circles. No extinction correction is applied to the magnitudes. 
Two different isochrones are overlaid (their parameters are shown above each panel) and are reddened with 
the cited values. The two variable candidates \#359 and \#834 and the two reference 
stars \#672 and \#851 are marked with empty circle and empty squares, respectively. 
}\label{cmd_optic}
\end{figure} 

We plotted all 273 stars with three and more epochs in $V$ band on CMD $V$ vs. $V-I$ (see Fig.\,\ref{cmd_optic}). Here, we used $VI$ magnitudes 
from the first epoch. We selected the stars that lie within the determined cluster radius and one can see on the diagram 
that the cluster members mainly represent the lower part MS population. More massive stars have already evolved from the MS and 
the TO-point is well visible at $V \sim 20.8$ mag. Some RGB stars are also present. For comparison, we plotted 
isochrones of PARSEC v1.1 \citep{bress}, which are reddened with the proper extinction (see Fig. \ref{cmd_optic}) to match the 
observed MS, TO-point, RGB and the clump. The used parameters are in consistence with the previously obtained ones.
The two candidate variables (empty circles in the latter figure) lie in the middle and in the upper part of the MS but their
luminosities are not high enough to suspect them neither as $\delta$ Scuti nor as even dimmer $\gamma$ Doradus pulsating variables. 
According to the best fitting isochrones the masses of both stars are close to 1 solar mass and the detected amplitude variations 
about 2 tenths of magnitude within a time span of a month may be related to stellar eclipses by a companion star -- a likely 
possibility, taking into account their position on the CMD. Another option is a rotational modulation by Sun-like spots -- the
variation study \citep{rad98} among a sample of Sun-like stars reveals at least one object - HD 129333 -- with similar amplitude.
However, due to the sparsity of the available data any further clarification of the nature
of the variability of these stars is not possible.

\section{Conclusions}
In an attempt to strengthen our knowledge of Whiting\,1 we used NIR photometry and determined the physical parameters of Whiting\,1. 
We fitted isochrones to $K$ vs. $(J-K)$ color-magnitude diagram and derived an age t=5.7$\pm$0.3\,Gyr, metallicity $z$=0.006$\pm$0.001 
([Fe/H]=$-$0.5$\pm$0.1) and distance modulus $(m-M)_0$=17.48$\pm$0.10. Our results
confirm that Whiting 1 is a young and moderately metal-rich globular cluster. It is one of the
youngest from the Sgr dSph.

In the NIR CMD, we noticed a clustering of objects at $\alpha_{J2000}=02^{h}02^{m}56.6^{s}$, 
$\delta_{J2000}=-03^{\circ}16\arcmin09\arcsec$ with radius r$\sim$1.5$\arcmin$, presumably a background 
galaxy cluster. We estimated a redshift and z$\sim$1 is based 
on a comparison with the colors of GALEV evolutionary synthesis models \citep{kot09} for E/Sa galaxies.

Our multi-epoch $V$ band monitoring led up to the discovery of two candidate variable stars within the 
cluster radius. Their differential light curves give maximum peak-to-peak variation of $\sim$0.2 mag for both 
candidates over the monitoring period of 35 days and taking in to account their position on the CMD diagram the stars are 
likely to be eclipsing binaries.

\section*{Acknowledgments}

Authors would like to thank Venelin Kozhuharov for his valuable help on fitting analysis and the anonymous referee for his/her 
constructive comments and suggestions that helped us to greatly improve to the paper.
The data used in the publication were retrieved from the ESO Science
Archive. They were obtained under program IDs: 59.A-9002(A), 
59.A-9908(A), and 60.A-9013(A). This publication makes use of data products from the Two Micron All Sky Survey, 
which is a joint project of the University of Massachusetts and the Infrared Processing and Analysis Center/California Institute 
of Technology, funded by the National Aeronautics and Space Administration and the National Science Foundation.

\bsp

\label{lastpage}

\end{document}